\documentclass[aps,prl,twocolumn,longbibliography,floatfix]{revtex4-2}
\usepackage{listings}
\usepackage[T1]{fontenc}
\usepackage[utf8]{inputenc}
\usepackage{newtxtext}
\usepackage{amsmath}
\usepackage{mathtools}
\usepackage{amsfonts}
\usepackage{bm}
\usepackage{amssymb}

\usepackage{newtxmath}
\usepackage[thinlines]{easytable}
\usepackage{booktabs}
\usepackage{color}
\usepackage{xcolor}
\definecolor{blue}{rgb}{0.2, 0.3, 0.85}
\definecolor{red}{rgb}{0.2, 0.3, 0.85}
\definecolor{darkgreen}{rgb}{0.0, 0.5, 0.0}
\usepackage{multibib}
\usepackage{todonotes}
\usepackage{color}
\usepackage{xcolor}
\definecolor{blue}{rgb}{0.25, 0.3, 0.95}
\definecolor{red}{rgb}{1,0,0}
\definecolor{darkgreen}{rgb}{0.0, 0.5, 0.0}

\usepackage{braket}

\usepackage[caption=false]{subfig}

\usepackage{graphicx} 

\newcommand{\typ}{\mathrm{typ}}

\usepackage{tikz}
\usepackage{textcomp}

\def\be{\begin{equation}}
\def\ee{\end{equation}}
\def\bea{\begin{eqnarray}}
\def\eea{\end{eqnarray}}

\usepackage[colorlinks,bookmarks=false,citecolor=blue,linkcolor=blue,urlcolor=blue]{hyperref}

\begin{document}

\title{Interaction-Driven Instabilities in the Random-Field XXZ Chain}
\author{Jeanne Colbois}
\affiliation{Laboratoire de Physique Th\'{e}orique, Universit\'{e} de Toulouse, CNRS, UPS, France}
\author{Fabien Alet}
\author{Nicolas Laflorencie}
\affiliation{Laboratoire de Physique Th\'{e}orique, Universit\'{e} de Toulouse, CNRS, UPS, France}
\date{\today}

\begin{abstract}
Despite enormous efforts devoted to the study of the many-body localization (MBL) phenomenon, the nature of the high-energy behavior of the Heisenberg spin chain in a strong random magnetic field is lacking consensus. Here, we take a step back by exploring the weak interaction limit starting from the Anderson localized insulator. 
Through shift-invert diagonalization, we find that below a certain disorder threshold  $h^*$, weak interactions necessarily lead to ergodic instability, whereas at strong disorder the AL insulator directly turns into MBL. 
This agrees with a simple interpretation of the avalanche theory for restoration of ergodicity. 
We further map the phase diagram for the generic XXZ model in the disorder $h$ -- interaction $\Delta$ plane. 
Taking advantage of the total magnetization conservation, our results unveil the remarkable behavior of the spin-spin correlation functions: in the regime indicated as MBL by standard observables, their exponential decay undergoes a unique inversion of orientation $\xi_z>\xi_x$. 
We find that the longitudinal length $\xi_z$ is a key quantity for capturing ergodic instabilities, as it increases with system size near the thermal phase, in sharp contrast to its transverse counterpart $\xi_x$.   \end{abstract}
\maketitle

\noindent\textbf{\textit{Introduction.}}
Understanding the subtle interplay between disorder and interactions in quantum systems is a major challenge in condensed matter physics. 
In particular, the many-body localization (MBL) problem remains a disputed issue, despite almost two decades of study~\cite{altshuler_quasiparticle_1997,jacquod_emergence_1997,gornyi_interacting_2005,basko_metalinsulator_2006,znidaric_many-body_2008,pal_many-body_2010,bardarson_unbounded_2012,nandkishore_many-body_2015,imbrie_many-body_2016,abanin_recent_2017,alet_many-body_2018,abanin_many-body_2019,sierant_2024}. 
Key debates revolve around the (existence of an) ergodicity-breaking transition at high energies, its possible universality class and associated finite-size behavior~\cite{luitz_many-body_2015,chandran_finite_2015,monthus_many-body-localization_2016,khemani_critical_2017,dumitrescu_kosterlitz-thouless_2019,goremykina_analytically_2019,morningstar_renormalization-group_2019,morningstar_many-body_2020,schiro_toy_2020,laflorencie_chain_2020,garcia-mata_critical_2022}, as well as the microscopic mechanisms driving the restoration of ergodicity~\cite{potter_universal_2015,vosk_theory_2015,de_roeck_stability_2017,thiery_many-body_2018,roy_percolation_2019,roy_fock-space_2020,Ha_many-body_2023}. 
In this context, it is instructive to first recall the original concern of the field~\cite{fleishman_interactions_1980,gornyi_interacting_2005,basko_metalinsulator_2006}: what is the fate of the Anderson localized (AL) insulator against weak interactions\,? 
Although this question is well posed, most numerical work has focused instead on the strongly interacting random-field Heisenberg chain (RFHC)~\cite{pal_many-body_2010,luca_ergodicity_2013,luitz_many-body_2015,gray_many-body_2017,doggen_many-body_2018,mace_multifractal_2019,sierant_polynomially_2020,laflorencie_chain_2020,chanda_time_2020,sierant_thouless_2020,abanin_distinguishing_2021,morningstar_avalanches_2022}, with few exceptions
\cite{pekker_hilbert-glass_2014,kjall_many-body_2014,sahay_emergent_2021,moudgalya_perturbative_2020,leblond_universality_2021,wahl_local_2022,laflorencie_topological_2022,biroli_large-deviation_2023}.
Over the past decade, the RFHC has gone from being the standard model of MBL to an increasingly controversial topic, primarily due to finite-size effects in the numerics~\cite{panda_can_2020,suntajs_ergodicity_2020,sierant_thouless_2020,sierant_polynomially_2020,abanin_distinguishing_2021,morningstar_avalanches_2022}, but also as very slow dynamics has been observed~\cite{doggen_many-body_2018,weiner_slow_2019,chanda_time_2020,PhysRevLett.124.243601,luitz_absence_2020,sierant_challenges_2022} in a regime previously thought to be deeply localized. 
This then led to very different conclusions, such as the absence of a genuine MBL phase~\cite{suntajs_quantum_2020,sels_dynamical_2021}, regardless of mathematical arguments~\cite{imbrie_diagonalization_2016,imbrie_many-body_2016} in favor of its existence in a related model, and clear signatures of non-thermal behavior in  various related experiments~\cite{schreiber_observation_2015,smith_many-body_2016,luschen_observation_2017,roushan_spectroscopic_2017,lukin_probing_2019,rispoli_quantum_2019,guo_observation_2021,leonard_probing_2023}.

\begin{figure}[t!]
    \centering
    \includegraphics[width=\columnwidth]{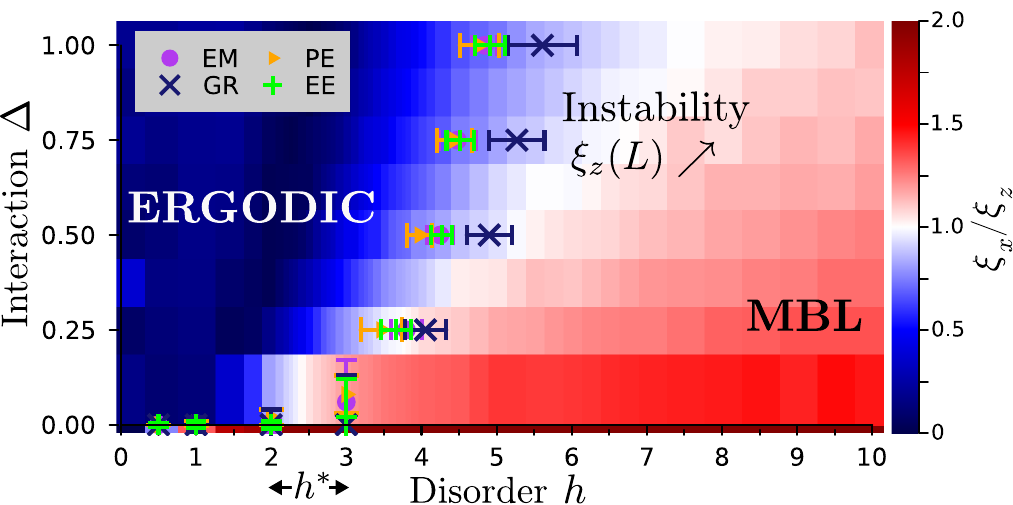}
    \caption{Disorder ($h$) -- interaction ($\Delta$) phase diagram of the XXZ chain Hamiltonian Eq.~\eqref{eq:H} at high energy (middle of many-body spectrum $\epsilon = 0.5$~\cite{luitz_many-body_2015}) 
in the $S^{z}_{\rm tot} = 0$ sector ($1/2$ for odd sizes). Symbols indicate the ergodic to MBL transition points obtained from standard observables (see legend and main text) by extrapolation of the crossings (see also Figs.~\ref{fig:Vscans} and~\ref{fig:xi},~\cite{sm}). 
    The  transition line starts from the Anderson insulator ($\Delta=0$) at a finite disorder strength $h^{*}$.
    At stronger disorder, the heatmap shows the ratio of the transverse to longitudinal mid-chain correlation lengths, fitted on sizes $L = 10, 12,14,16$, starting with $\xi_x/\xi_z \approx 2$ at $\Delta =0$. The region where $\xi_x/\xi_z\lesssim 1$ gives a rough approximation of the instability regime in which 
    $\xi_z(L)$ increases with $L$ (see also  Fig.~\ref{fig:xi} and main text for a discussion of this instability).
     \label{fig:PD}}
\end{figure}    

\vskip 0.15cm
\noindent\textbf{\textit{Main results.}}
In this Letter, we consider the XXZ spin chain
\begin{equation}
    {\cal{H}}_{\Delta}=\sum\nolimits_{i}\left(S_i^x S_{i+1}^{x} + S_i^y S_{i+1}^{y} + \Delta S_i^z S_{i+1}^{z}+h_iS_i^z\right),\label{eq:H}
\end{equation}
in its easy-plane version $\Delta\in [0,1]$ ($\Delta=1$ being the RFHC), and random fields uniformly drawn in $[-h,\,h]$. ${\cal{H}}_{\Delta}$ conserves $S^z_{\rm{tot}} =\sum_i S_i^z$, which will be highly exploited in this work. 
The high-energy phase diagram of this model is shown in Fig.~\ref{fig:PD}, building on mid-spectrum shift-invert diagonalization~\cite{luitz_many-body_2015,pietracaprina_shift-invert_2018} of periodic chains of sizes $L \in [10 \dots 21]$. Before giving a detailed description, we first sketch our main findings.
(i) Along the non-interacting line $\Delta=0$, we argue, and numerically verify, that there is a disorder threshold $h^*$ for the AL insulator below which any finite interaction restores ergodicity, while at strong disorder, AL turns into MBL. 
(ii) For finite $\Delta$, we provide an extrapolation ($L\to \infty$) of the MBL transition line using standard estimates (Fig.~\ref{fig:PD}, gray region).
(iii) Spin-spin correlation functions (longitudinal $zz$ and transverse $xx$, with respect to the random field) and their associated correlation lengths $\xi_{x,\,z}$ present remarkably contrasted behaviors on the MBL side: the dominant orientation $\xi_x>\xi_z$ (inherited from AL) changes to $\xi_x<\xi_z$ across the phase diagram (see color map in Fig.~\ref{fig:PD}).
At intermediate disorder, this is accompanied by a growth of $\xi_z$ with $L$, while $\xi_x$ barely changes. We interpret these observations as a qualitative sign of increasing ergodic instabilities, based on the different behavior of these correlators in the ergodic phase.
(iv) For larger $h$, both AL and MBL regimes show very short and size-independent $\xi_{z,x}\ll 1$,
suggesting that an ergodic instability is very unlikely.

\newpage\noindent\textbf{\textit{Ergodic instability at weak interactions.}}
The most discussed theoretical framework describing the restoration of ergodicity from the MBL phase is provided by the avalanche theory (AT)~\cite{de_roeck_stability_2017,thiery_many-body_2018}.
In a nutshell, the AT starts from strong disorder and considers a possible runaway delocalizing instability triggered by rare ergodic seeds.
This mechanism is predicted to occur when the length scale $\zeta$, which controls the exponential decay of the effective coupling between the ergodic bubble and the surrounding localized spins, exceeds a certain ${\cal{O}}(1)$ critical threshold $\zeta_{\rm avl.}$~\cite{noteAT}. A good starting point for estimating $\zeta$ is the non-interacting AL limit, since this length scale can be identified~\cite{crowley_avalanche_2020} by the exponentially localized Anderson orbitals, controlled by the disorder-dependent average AL length~\cite{colbois_breaking_2023} $\xi_{\rm AL}=1/{\ln\left[1+\left({h}/{h_0}\right)^2\right]}$~\cite{noteAL}.
A naive application of AT in the vanishing interaction limit~\cite{crowley_avalanche_2020}, where $\zeta\sim \xi_{\rm AL}$, predicts the existence of a finite disorder strength below which an instability condition is met. Since the localization length typically increases with the interaction strength~\cite{shepelyansky_coherent_1994,frahm_effective_1996,weinmann_thouless_1997}, an interaction-driven ergodic instability seems inevitable below a certain threshold disorder strength $h^*$. Conversely, at higher disorder, a completely different scenario emerges in which the localized regime remains stable against weak interactions. These simple ideas will be compared with numerical simulations below.
\begin{figure}[b!]
    \includegraphics[width=.95\columnwidth]{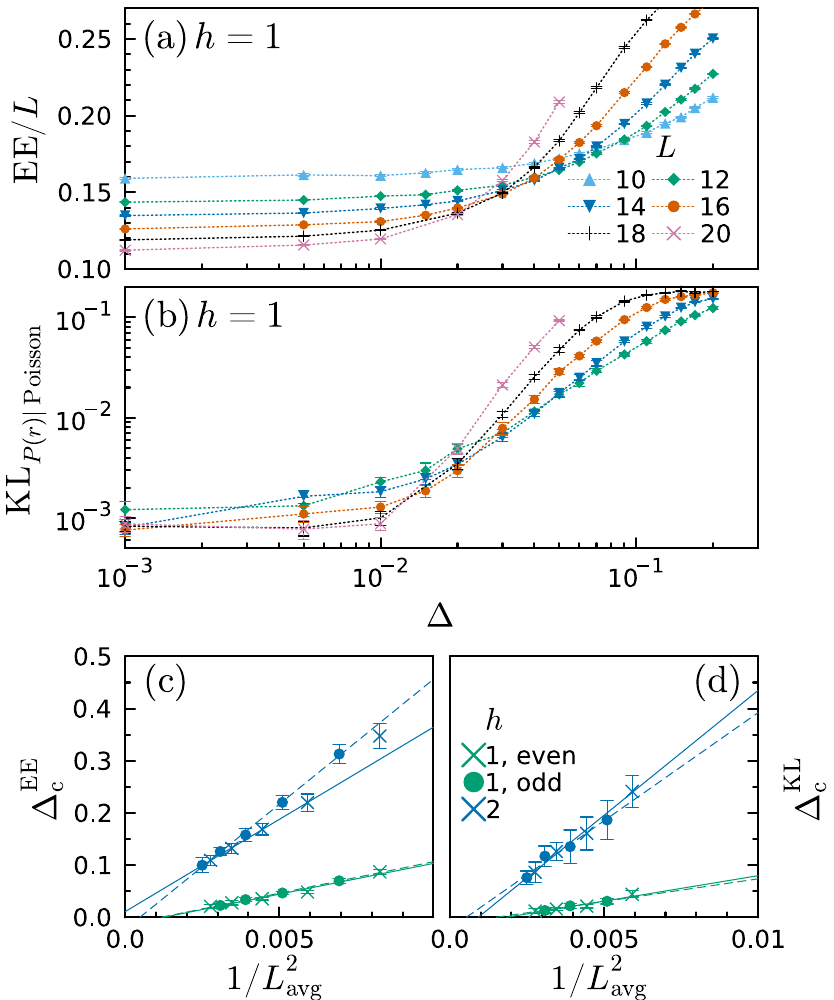}
\caption{\label{fig:Vscans} Constant fields $h=1$ and $h=2$ data: critical interaction strength $\Delta_c$ above which ergodicity occurs, as estimated from mid-chain entanglement entropy (EE) and gap ratio (GR) statistics. At finite sizes, upon decreasing the interaction $\Delta$, (a) EE changes from a volume-law at large $\Delta$ to an area law at small $\Delta$, and (b) the Kullback-Leibler divergence between the gap ratio distribution and the Poisson statistics goes from $\approx 0.1895$ to $0$ (in practice, to a finite value set by the precision of the numerical distribution)~\cite{sm}. Panels (c) and (d) show the drifts of the critical interaction strength from both EE and GR crossings of even (crosses, solid fits) and odd (circle, dashed fits) sizes, with $L_{\rm avg}$ indicating the mean of the two sizes. The best fits yield a quadratic dependence with $1/L_{\rm avg}$.}
\end{figure}

\vskip 0.15cm

\noindent\textbf{\textit{Phase diagram from standard observables.}}
Using shift-invert diagonalization of finite chains, we obtain mid-spectrum eigenstates of Eq.~\eqref{eq:H}, and first verify the existence of an ergodic instability at weak disorder in the small $\Delta$ limit. Fig.~\ref{fig:Vscans} displays results for vertical scans at $h=1,\,2$ (where $\xi_{\rm AL}\approx 1.9,\,0.7$).  
The top panels display two standard markers for the transition: mid-chain entanglement entropy (EE), and a measure of the spectral statistics based on the Kullback-Leibler divergence between the gap ratio distribution and the Poisson statistics ${\rm{KL}}_{[P(r)|{\rm{Poisson}}]}$~\cite{sm}. 
The bottom panels demonstrate a very clear drift of the crossing positions for even and odd sizes with increasing system lengths, thus showing that $\Delta_c(L) \to 0$ for $h=1$ and $h=2$. For stronger disorder, the system is in a finite-size crossover regime, yielding very large errors and a very strong drift for $\Delta_c(L)$ at $h=3$, making the extrapolation difficult~\cite{sm} (see Fig.~\ref{fig:PD}). 
Although this makes it challenging to extract a precise numerical value for the non-interacting threshold $h^*$, our data clearly point to a finite and not too small value $2\leq h^{*}\leq3$, corresponding to a rather short AL length $0.7\ge \xi_{\rm AL}^{*}\ge 0.5$. 

In Fig.~\ref{fig:PD}, we report this ergodic instability of the Anderson insulator at vanishing interaction strength, using two further measures: the eigenstate ergodicity in the Hilbert space quantified by the participation entropy (PE)~\cite{luca_ergodicity_2013,luitz_many-body_2015,mace_multifractal_2019,pietracaprina_hilbert-space_2021}, and  the extreme statistics of the local magnetizations (EM)~\cite{laflorencie_chain_2020,colbois_breaking_2023}. 
These results bring insight to the observations in Ref.~\cite{lin_many-body_2018}, as well as into the results of Ref.~\cite{leblond_universality_2021}, where the reported onset of quantum chaos is indeed naturally explained by the fact that the study was performed in at low disorder  ($h\approx0.57, \,\xi\approx 5$). 
A similar scenario occurs in the disordered interacting Majorana chain, where an immediate ergodic instability arises when the non-interacting localization length exceeds a threshold~\cite{laflorencie_topological_2022}.

\begin{figure*}[t!]
    \includegraphics[width=2\columnwidth]{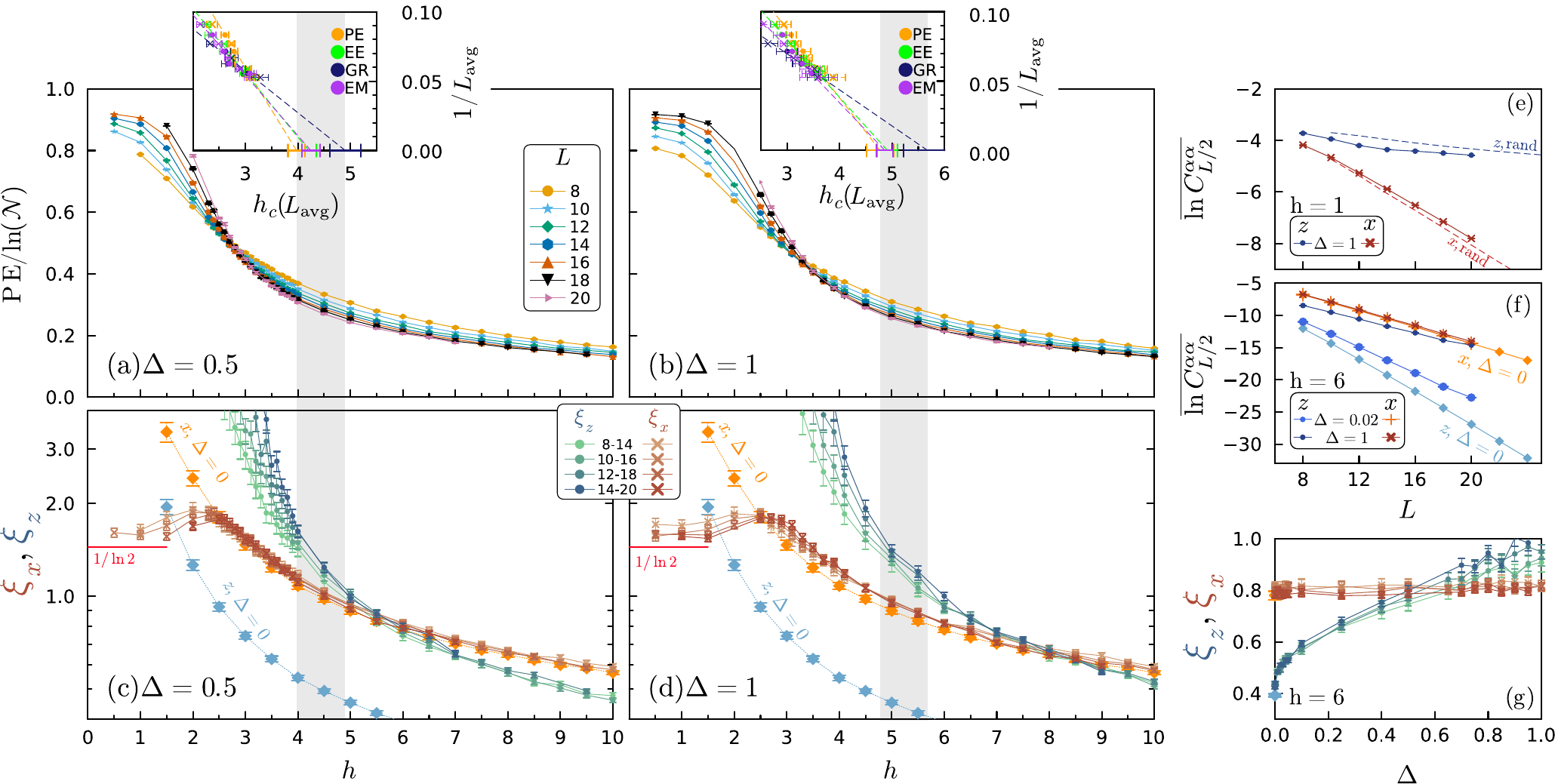}
\caption{Overview of the ergodic, instable and MBL regimes, a function of the disorder $h$ and the interaction $\Delta$. In the four left panels, the gray area indicates the transition extrapolated from four standard observables (insets). (a) and (b) At $
\Delta=0.5$ and $\Delta = 1$, the PE scaling with the logarithm of the Hilbert space dimension  $D_1 \ln{\cal{N}} + b_1$ shows a crossing point which corresponds to a change of sign for the subleading correction $b_1$.
(c) to (g) Mid-chain correlations and corresponding correlation lengths. (c) and (d) Mid-chain correlation lengths at $\Delta = 0.5$ and $\Delta = 1$ compared to the AL case, extracted from fits on even sizes of the typical value of the mid-chain connected correlations Eq.~\eqref{eq:C}.
Right panels show the associated characteristic decay: (e) at $h = 1, \Delta = 1$ in the ergodic regime, compared to the correlations for a random vector (dashed lines); (f) at $h = 6, \Delta = 0.02$ in the MBL regime and at $h = 6, \Delta = 1$ in the regime showing instabilities, compared to the correlations in the Anderson case for the same disorder strength.
(g) As a function of the interaction strength at $h = 6$, the longitudinal correlation length shows a square-root like behavior. In all cases, the AL values are extracted from fits on $L= 14 \cdots 20$.
\label{fig:xi}}
\end{figure*}

Moving away for the AL line, we investigate the extent of the ergodic regime at finite interaction, building on the same four standard observables. The top panels of Fig.~\ref{fig:xi} report the behavior of PE for various sizes as a function of the disorder $h$ for fixed $\Delta = 0.5$ and $\Delta=1$ (RFHC).
In agreement with previous observations, the ergodic regime is clearly underestimated by small system sizes, leading to  systematic finite-size drifts for the ergodic-to-MBL crossovers towards larger disorder strengths~\cite{sm}, as clearly shown in the top insets of Fig.~\ref{fig:xi}.
However, using linear fits with $1/L$ as in the RFHC case~\cite{sierant_polynomially_2020}, 
we find that these drifts all consistently converge to finite values $h_c$, clearly indicating a broad but finite extent of the ergodic regime, as shown by the various symbols in Fig.~\ref{fig:PD}.

It should be noted that the extrapolated critical disorder obtained from the spectral statistics transition $h_{c}^{\rm GR}$ systematically yield the largest critical disorder strengths. For $\Delta =1$, our estimate $h_{c}^{\rm GR}=5.6(5)$ is consistent with the numerical landmark $h_{\rm mg}\approx 5.7$ proposed by Morningstar {\it{et al.}}~\cite{morningstar_avalanches_2022}, where the smallest gap starts to deviate from Poisson expectations.

\noindent\textbf{\textit{New insights from spin-spin correlations.}}
In order to provide a real-space view of how  ergodicity is reached from the strong disorder regime, we now focus on spin correlations for mid-spectrum eigenstates of ${\cal{H}}_{\Delta}$. Although introduced early on as possible indicators of the MBL transition for the RFHC~\cite{pal_many-body_2010}, they have been the subject of surprisingly little work for U(1) symmetric models~\cite{pal_many-body_2010,lim_many-body_2016,colmenarez_statistics_2019,villalonga_characterizing_2020,Kulshreshtha_approximating_2019,varma_length_2019}, leading nonetheless to the observation of their non-trivial distributions in both ETH and MBL regimes~\cite{colmenarez_statistics_2019} or that they may signal an intermediate critical phase driven by rare events~\cite{lim_many-body_2016}.

Here we show that pairwise spin correlation functions provide remarkable insights, for several reasons (see also Ref.~\cite{colbois_companion_2024} for further results and detail). First, exponentially decaying correlators allow one to connect to the non-interacting Anderson limit, where the localization and spin correlation lengths are all proportional (at least for not too weak disorder, typically above $ h \sim 2$). Moreover, correlation lengths remain well-defined in the MBL regime at strong disorder. 
Finally, considering correlations at the largest possible distance in periodic chains can probe early onsets of thermalization processes through system-wide resonances. For this, we examine the mid-chain connected correlation functions~\cite{abs} 
\be
C_{L/2}^{\alpha\alpha}=\left|\langle S_i^{\alpha} S_{i+L/2}^{\alpha}\rangle - \langle S_i^{\alpha} \rangle\langle S_{i+L/2}^{\alpha}\rangle\right|.
\label{eq:C}
\ee
Fig.~\ref{fig:xi} (e-f)  illustrates the two limiting cases for the characteristic decays with $L$ of the typical average of the longitudinal ($\alpha=z$) and transverse ($\alpha=x$) components.
In the ergodic phase, where the eigenstate thermalization hypothesis (ETH) applies~\cite{deutsch_quantum_1991,srednicki_chaos_1994,rigol_thermalization_2008}, eigenstates are well described by featureless random states, which yield an absence of spatial variation for both components $x$ and $z$~\cite{colbois_companion_2024}. 
However, we expect strongly contrasted dependences on the length $L$: while ETH implies an exponential vanishing with the system size for $C^{xx}_{L/2}\sim {{\cal{N}}}^{-1/2}$
(with ${\cal{N}}\approx 2^{L}$ the size of the Hilbert space), the total magnetization conservation ($S^{z}_{\rm tot}={\rm{constant}}$) leads to a much slower algebraic decay of the longitudinal component $C^{zz}_{L/2}\sim ({4L})^{-1}$. In Fig.~\ref{fig:xi} (e), ED results deep in the ergodic regime reproduce such very different behaviors, approaching the exact calculation performed with $S^{z}_{\rm tot}=0$ random states.

Conversely, although the magnetization conservation still holds, the strong disorder regime shows exponential decays for both $x$ and $z$ components, as shown for $h=6$ and weak enough interaction in Fig.~\ref{fig:xi}(f), see also panel (g). It is therefore natural to define the typical mid-chain correlations lengths $\xi_x$ and $\xi_z$~\cite{note_xi} as follows
\be
{\overline{\ln C^{\alpha\alpha}_{L/2}}} =:-\frac{L}{2\xi_{\alpha}}+{\cal{O}}(1),
\ee
where ${\overline{(\ldots)}}$ stands for disorder averaging.
We first make a few key observations for the large-$h$, presumably localized regime.

(i) In the non-interacting AL limit ($\Delta=0$) the $z$-component of the spins are pinned by strong random fields, yielding rather short correlation lengths and dominant quantum fluctuations in the transverse channel. As a result, we observe at $h=6$, $\xi_x^{\rm{AL}} = 0.80(2) > \xi_z^{\rm{AL}} = 0.41(1)$. This factor of 2 appears to be robust along almost the entire AL line, at least for $h> 2$ (see Fig.~\ref{fig:PD}), providing a hallmark for this regime.

(ii) Remarkably, even a very weak interaction above the AL line leads to a strong qualitative change for the longitudinal correlations with a sharp increase of the correlation length $\xi_z$, while the transverse part remains essentially unaffected, as clearly visible in Fig.~\ref{fig:xi}(f-g).
In particular, we observe in Fig.~\ref{fig:xi}(g) a very rapid, square-root-like increase of $\xi_z$ with $\Delta$, while $\xi_x$ remains approximately at its non-interacting value.

(iii) For larger interactions, the trend has strongly intensified so that $\xi_z$ crosses $\xi_x$ around $\Delta\sim 0.5$, and eventually becomes larger as illustrated by the Heisenberg point at $\Delta = 1, h=6$. 

\vskip 0.15cm
\noindent\textbf{\textit{Interaction-driven instabilities.}}
This striking reversal in the orientation of the dominant correlations, already visible at small size in the color plot provided over the entire phase diagram, is even more apparent with larger $L$ in Fig.~\ref{fig:xi} (c-d) where one observes crossing between $\xi_x$ and $\xi_z$ for two representative horizontal cuts ($\Delta=0.5,\, 1$). 
In addition, both plots show that the transverse correlation length $\xi_x$ has a very weak size dependence and remains finite and small everywhere. The only significant interaction effect appears in the ergodic regime below $h \approx 2.5$, where $\xi_x$ deviates from the divergence of $\xi_x^{\rm AL}$ and decays to its ETH value $1/\ln 2$. Quite differently, Fig.~\ref{fig:xi}(c-d and g)  show that the longitudinal correlation length $\xi_z$ is much more sensitive and displays very pronounced variations both with $\Delta$ (in all regimes) and with $L$ (in the ergodic phase). 
The latter  is easily understood from the distinction between the algebraic and exponential decay of the typical $C^{zz}_{L/2}$ {\it{vs.}} $C^{xx}_{L/2}$ in the ergodic regime, where effectively $\xi_x/\xi_z\to 0$ as $L$ increases. 

In the opposite case of strong disorder and weak interaction, the situation is qualitatively very close to AL, with the observed hierarchy $\xi_x>\xi_z$. This suggests that MBL and AL are connected in this part of the phase diagram~\cite{de_tomasi_efficiently_2019,colbois_breaking_2023}, where furthermore no finite-size effects are observed for either $\xi_z$ or $\xi_x$. 
However, as the disorder is gradually reduced, though still well before the presumed ergodic transition, an instability is observed via a systematic growth with $L$ of our numerical estimates of $\xi_z$. 
This is clearly visible in Fig.~\ref{fig:xi} where one sees such instabilities, for example as soon as $h< 7$ for $\Delta=0.5$ (panel c) or above $\Delta\approx 0.1$ for $h=6$ (panel g), both cases corresponding to a regime where all other standard finite-size observables  show well-converged MBL-like behavior. This striking finding is a stronger signal than the observed $\xi_x=\xi_z$ crossing, even though it occurs in roughly the same regime. Indeed, such a crossing simply reflects the fact that $\xi_z$ increases faster than $\xi_x$, but the additional growth of $\xi_z$ with $L$ is a remarkable indication of an anomalous response of the diagonal correlations in a regime of disorder where one would have rather expected MBL physics. 
Based on the very different scaling with $L$ of $C_{L/2}^{zz}$ in the two opposite regimes, we may therefore interpret these observations as a qualitative marker of emerging ergodic instabilities in models conserving the total magnetization.

\vskip 0.15cm
\noindent\textbf{\textit{Summary and discussion.}}
In this Letter, we first provided evidence for the direct instability of the Anderson insulator towards ergodic behavior in the small interaction, weak disorder limit, $h<h^*\sim  2-3$. 
This prediction is directly testable experimentally in platforms where interactions can be controlled, such as with Feschbach-resonances in cold-atoms~\cite{chin_feshbach_2010,schreiber_observation_2015}. 
In a second part, we took advantage of the magnetization / particle number conservation (often met in experiments) to unveil a finite-size growing instability of $\xi_z(L)$, in a part of the phase diagram where other finite-size indicators point to an MBL regime. 
Our main results, summarized in Fig.~\ref{fig:PD}, will be further detailed and expanded in a forthcoming companion paper~\cite{colbois_companion_2024}.
The system-wide response probed by $C^{zz}_{L/2}$ and its two-body nature can be linked  to the end-to-end mutual information~\cite{tomasi_quantum_2017}, which was recently used as a landmark to detect system-wide resonances~\cite{morningstar_avalanches_2022}. 
Two remarks are in order though: by averaging over eigenstates and disorder realizations, $\ln C^{zz}_{L/2}$ reflects the typical behavior whereas Ref.~\cite{morningstar_avalanches_2022} studied the extreme statistics of the maximal (over all eigenstates) mutual information. 
Second, the connected correlator $C^{zz}$ is routinely accessible as a density-density correlation in most MBL experimental platforms, see e.g. Ref.~\cite{lukin_probing_2019}, which indeed observe an increase of $\xi_z$ as disorder is decreased in the MBL regime, albeit in a different setup than ours (correlations after a quench in a quasi-periodic potential versus correlators in eigenstates in a random potential). 

Could the growing $\xi_z$ with $L$ be further considered as a smoking gun of avalanches\,? Our eigenstates analysis does not allow to conclude on this, but we remark that ergodic inclusions (potential seeds for avalanches) naturally favor an enhancement of diagonal correlations.  Indeed,  typical $zz$ correlations barely decay across ergodic regions, yielding an effective growth of $\xi_z$, that may be a key quantity to further investigate avalanche instabilities in realistic microscopic models (for a recent study of dynamical correlations in this context, see~\cite{szoldra_catching_2024}).
Nevertheless, one of the most difficult remaining questions concerns the status of the intermediate region showing finite-size instabilities.Will the growth of $\xi_z(L)$ continue on larger length scales, giving rise to an ergodic regime, or a novel intervening glassy state~\cite{biroli_large-deviation_2023}, or will it instead saturate, corresponding to an MBL phase\,? We hope that our work will motivate further explorations in particle number-conserving theoretical models or in experimental platforms susceptible to host an MBL transition.

\acknowledgements  \vskip 0.15cm
\noindent\textbf{\textit{Acknowledgements.}}
We are glad to acknowledge inspiring discussions with G. Biroli,  T. Giamarchi,  F. Heidrich-Meisner, T. Prosen, M. Tarzia,  R. Vasseur.
This work has been partly supported by the EUR grant NanoX No. ANR-17-EURE-0009 in the framework of the ''Programme des Investissements d'Avenir'', is part of HQI initiative (www.hqi.fr) and is supported by France 2030 under the French National Research Agency award number ANR- 22-PNCQ-0002, and also benefited from the support of the Fondation Simone et Cino Del Duca. 
We acknowledge the use of HPC resources from CALMIP (grants 2022-P0677 and 2023-P0677) and GENCI (projects A0130500225 and A0150500225), as well as of the PETSc~\cite{petsc-user-ref,petsc-efficient}, SLEPc~\cite{slepc-toms,slepc-manual}, MUMPS~\cite{MUMPS1, MUMPS2} and Strumpack~\cite{Strumpack} sparse linear algebra libraries. 

\bibliography{mbl,notes}

\begin{thebibliography}{104}
\bibitem [{noteAT()}]{noteAT}%
  \BibitemOpen
  \bibinfo {note} {The precise ${\cal{O}}(1)$ value of $\zeta_{\rm avl.}$ is
  set by the many-body level spacing, and depends on the modelization of the
  coupling between the ergodic seed and the localized spins.}\BibitemShut
  {Stop}%
\end{thebibliography}

\vskip 1cm
\setcounter{section}{0}
\setcounter{secnumdepth}{3}
\setcounter{figure}{0}
\setcounter{equation}{0}
\renewcommand\thesection{S\arabic{section}}
\renewcommand\thefigure{S\arabic{figure}}
\renewcommand\theequation{S\arabic{equation}}

\newpage
\onecolumngrid
\begin{center}
    \bfseries\Large Supplemental material
\end{center}
    Here, we provide some details in support of the main text. In Sec.~\ref{sec:Obs}, we discuss two obervables : the Kullback-Leibler divergence of the gap ratio distribution, and the extreme magnetization. 
    In Sec.~\ref{sec:Crossings}, we give more details regarding the location of the crossing pointsand their drift. Finally, in Sec.~\ref{sec:Params}, we give the detail of the number of samples and eigenstates kept for each size.

    \maketitle
    
    \section{Observables}
    \label{sec:Obs}
    \subsection{Kullback-Leibler divergence of the gap ratio distributions}

    Perhaps the most standard observable to distinguish between chaotic and integrable systems is the level statistics~(see for instance \cite{rosenzweig_repulsion_1960,jacquod_emergence_1997,oganesyan_localization_2007,pal_many-body_2010,alet_many-body_2018,giraud_probing_2022} and references therein). 
    The gap ratio involves three consecutive energy levels and is defined as~\cite{oganesyan_localization_2007}
    \begin{equation}   
        r_i = \min\left(\frac{s_i}{s_{i-1}}, \frac{s_{i-1}}{s_{i}}\right),
    \end{equation}
    where $s_i = E_{i}-E_{i-1}$ is the spacing between two consecutive levels. 
    
    In the case of an integrable system, the gap ratio follows a distribution related to Poisson statistics:
    \begin{equation}
        P_{\rm{Poisson}}(r) = \frac{2}{(1+r)^2}.
        \label{eq:poisson}
    \end{equation}
    By contrast, in a system satisfying the eigenstate thermalization hypothesis (ETH), the distribution is given by the RMT spectrum in the GOE ensemble ~\cite{atas_distribution_2013}
    \begin{equation}
        P_{\rm{GOE}}(r) = \frac{27}{4} \frac{r+r^2}{(1+r+r^2)^{5/2}}. 
     \end{equation}

    \begin{figure}[h!]
        \includegraphics[width = 0.65\textwidth]{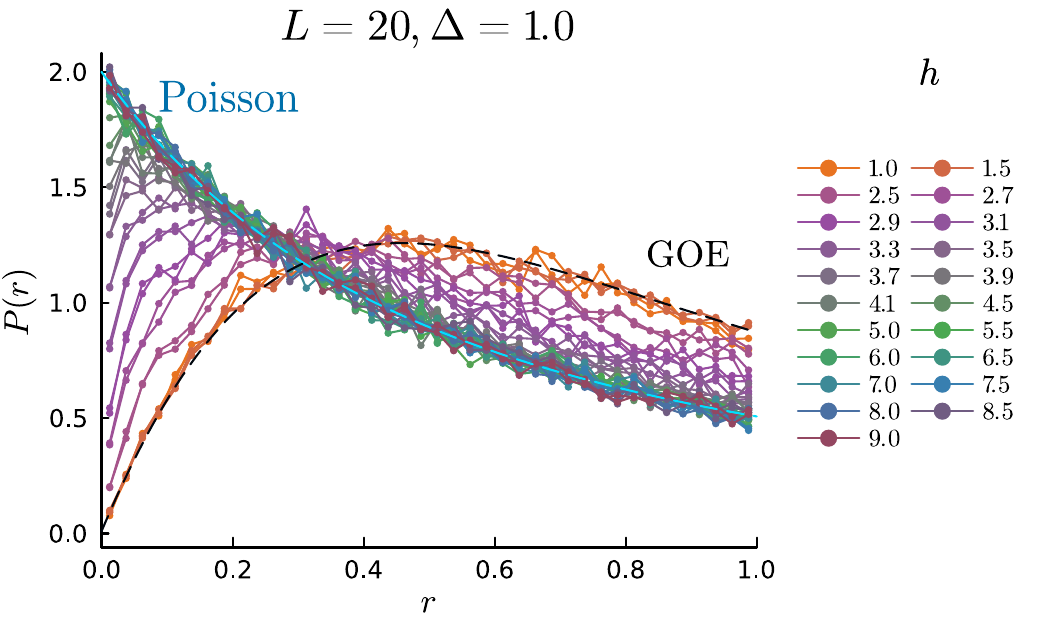}
        \caption{\label{fig:GR} Distribution of the gap ratios for $\Delta = 1$ for a system of size $L= 20$, with 41 bins and two different samplings. The dashed lines represent the Poisson and the GOE statistics.}
    \end{figure}
    
    As illustrated in Fig.~\ref{fig:GR}, the gap ratio distribution in the Heisenberg chain goes from $P_{\rm{GOE}}$ at weak disorder to $P_{\rm{Poisson}}$ at strong disorder.
    The standard approach to evaluate this qualitative change in a single quantitative number is to use the average gap ratio~\cite{pal_many-body_2010}
    $$r_{\rm avg}=\int_{0}^{1}rP(r){\rm{d}}r.$$ This is however numerically challenging to use in the limit of strong disorder or weak interactions because a lot of the weight of the integral is not useful to discriminate between the Poisson distribution and the numerical distribution.
    It was suggested to use instead the integral of the gap ratio distribution from $r=0$ to $r = 1/4$~\cite{giraud_probing_2022} in order to take advantage of the region where the two distributions differ the most, and to provide a probe that remains easy to evaluate experimentally. 
    In a similar spirit, taking advantage of the full distribution, an alternative would be to use the normalized integral of the difference between the distributions up to some value 
    $r_{\max}$~\cite{jacquod_emergence_1997}, 
    $$\eta = \frac{\int_0^{r_{\max}}\left(P(r)-P_{\rm{GOE}}(r)\right){\rm{d}}r}{\int_0^{r_{\max}}\left(P_{\rm{Poisson}}(r)-P_{\rm{GOE}}(r)\right){\rm{d}}r}.$$

    Here, with a similar aim of taking full advantage of the two distributions, and further trying to ensure a large dynamics close to the Poisson regime, we decided to use the Kullback-Leibler divergence~\cite{kullback_information_1951} between the numerical gap ratio distribution and the Poisson statistics from Eq.~\eqref{eq:poisson}. 
    We find that it is more stable numerically to compute the KL divergence of the latter with respect to the former, rather than the opposite:
    \begin{equation}
        {\rm{KL}}_{P | {\rm{Poisson}}} = \int_0^{1} P(r) \ln\left(\frac{P(r)}{P_{\rm{Poisson}}(r)}\right){\rm{d}}r.
        \label{eq:KL}
    \end{equation}
    The KL divergence therefore goes from ${\rm{KL}}_{P_{\rm{GOE}} | {\rm{Poisson}}} \approx 0.1895$ in the ETH regime, to zero in the localized regime. 
    In practice, the value is lower bounded by the numerical precision we can achieve based on the histograms obtained from our samples.
    The main challenge is to obtain errors on this gap ratio statistics. In practice, we do this computing 30 times the KL divergence for different samplings of the GR data (using $20$k ratios sampled randomly among the data if $L< 12$ and $30$k ratios otherwise), with two different number of bins. 
    Numerically, the form in Eq.~\eqref{eq:KL} ensure that an empty bin gets correctly assigned a weight zero. 
    The corresponding histograms are illustrated in Fig.~\ref{fig:GR} for $\Delta =1$.
    The results obtained from this procedure at constant field are shown in Fig.~2 of the main text. In Fig.~\ref{fig:crossings}, left panel, we display how the crossings behave at constant interaction $\Delta = 0.25$.
    
    \subsection{Extreme magnetization}
    \begin{figure}[h!]
        \includegraphics[width=\textwidth]{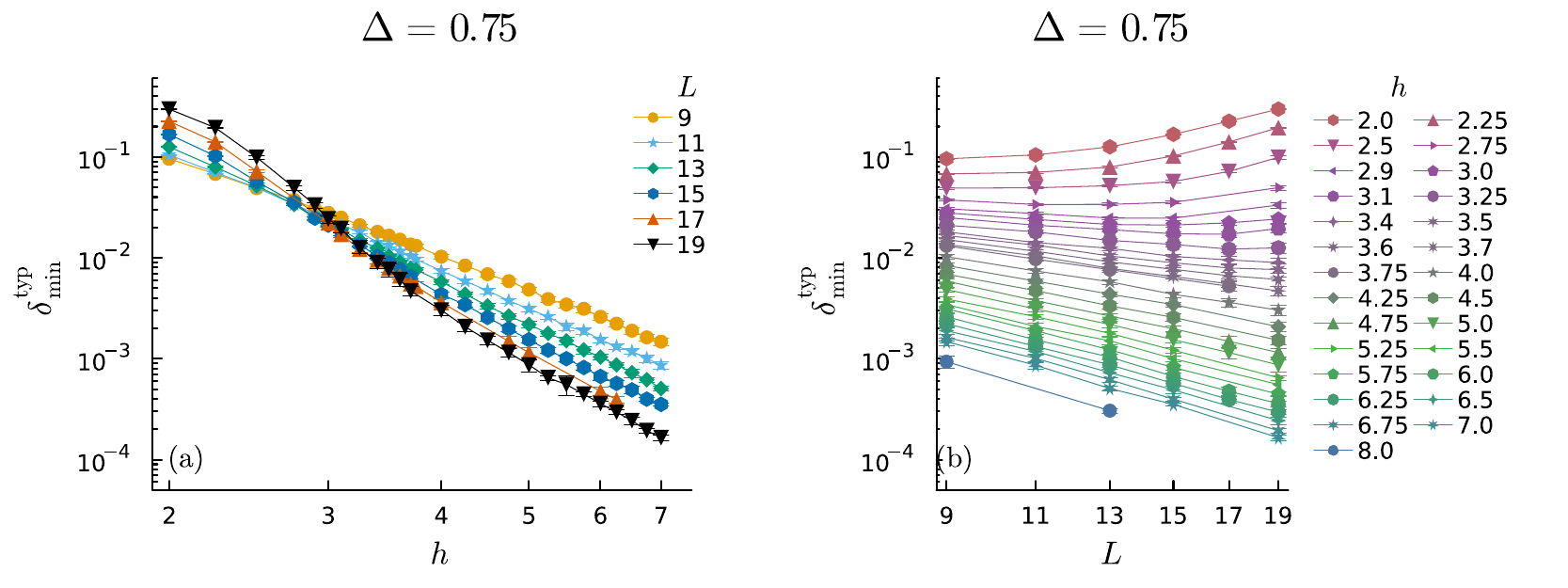}
        \caption{\label{fig:EMd075} Typical minimal deviations for odd sizes at $\Delta = 0.75$ (a) as a function of the disorder strength for various system sizes; (b) the same data presented as a function of the size, showing a clear power-law decay with system size as in Eq.~\eqref{eq:deltapow}.  }
    \end{figure}
    
    We use a simple probe for the most polarized (``most classical'') site in the chain : the extreme magnetization (EM), characterized by the minimal deviation $\delta_{\min}$~\cite{laflorencie_chain_2020,colbois_breaking_2023}
    For a given sample and eigenstate, the minimal deviation is given by 
    \begin{equation}
        \delta_{\min} = \min_{i} \left(1/2 - |\langle S_i^z \rangle|\right),
    \end{equation}
    where $i$ indexes the sites and where the expectation value is taken in the given eigenstate. 
    In the Anderson case, the typical minimal deviation goes to zero as a power law of the system size
    \begin{equation}
        \delta_{\min}^{\typ}  \sim L^{-\gamma_{\rm{AL}}(h)}
    \end{equation}
    where the power-law exponent 
    \begin{equation}
        \gamma_{\rm{AL}}(h) \approx \frac{1}{2\xi_{\rm{AL}} \ln 2}
        \label{eq:gammaal}
    \end{equation}
    is controlled by the (typical) Anderson localization length.
    
    Numerically, this behavior is also observed at strong enough disorder in the Heisenberg chain~\cite{laflorencie_chain_2020}, and in the XXZ chain as can be seen in Fig.~\ref{fig:EMd075} for $\Delta = 0.75$. This defines the disorder- (and interaction-) dependent exponent $\gamma$:
    \begin{equation}
        \delta_{\min}^{\typ}  \sim A L^{-\gamma}
        \label{eq:deltapow}
    \end{equation}
    
    In contrast, in the ergodic regime, the typical minimal deviation goes to 1/2, as imposed by the ETH, and for finite sizes the effective exponent can take negative values.
    It is expected that this exponent shows a jump at the transition~\cite{laflorencie_chain_2020} from its critical value $\gamma_c$ to zero, but the value of $\gamma_c$ is unknown, and it is unclear whether it is constant at the transition. 

    To determine the critical disorder strength, there are several possibilities. In Ref.~\cite{laflorencie_chain_2020}, a scaling approach was performed for the behavior of $\ln(\delta_{\min}^{\typ})$ vs $L$ (in the MBL regime) and vs the Hilbert space size $\mathcal{N}$ (in the ergodic regime) yielding the critical disorder strength $h_c(\Delta=1)\sim4.2(5)$.
    We note also that in Ref.~\cite{colbois_breaking_2023}, it was shown that the KL divergence between the minimal deviation distributions in the Heisenberg and the XX chains show a crossing that signals an extreme value transition, but it is yet to be clarified whether this is directly probing the MBL transition.
    In our systematic exploration of the phase diagram, we turn to a simpler approach. 
     Since it is a priori unclear whether $\gamma_c$ is constant accross the transition out of the ergodic regime, we can not use this value as a criterion.
    As an alternative, we use the crossings between the different sizes for $\delta_{\min}^{\typ}(h)$ as a criterion. 
    This roughly corresponds to looking for the change of sign in the effective $\gamma$, which means that for each finite size, the crossing obtained tends to underestimate the extent of the ergodic phase -- a challenge common to all standard probes in this model.
    However, we note that at $\Delta = 1$, the obtained critical value $h_{\mathrm{c}}^{\rm EM} = 4.8 \pm 0.16$ is slightly larger than the result of Ref.~\cite{laflorencie_chain_2020} (see Table~\ref{tab:extrapolated}). 
    We also observe (e.g. in Fig.~1 and~3 of the main text and in Fig.~\ref{fig:consth} of this supplemental material) that the resulting drift yields results compatible with other standard observables such as the EE. 
    
    \section{Drifts of the crossings and attempt at an extrapolation}
    \label{sec:Crossings}
    
    The results for the crossings at constant interactions are summarized in Table~\ref{tab:extrapolated}
    
    \begin{table}[ht]
        \centering
        \begin{tabular}{c|l|l|l|l}
            \toprule
            Observable & {$\Delta = 0.25$} & {$\Delta = 0.5$} & {$\Delta = 0.75$} & {$\Delta = 1$} \\
            \midrule
            PE & 3.44(28) & 3.98(16) & 4.44(23) & 4.77(26) \\
            EE & 3.67(20) & 4.21(14) & 4.50(18) & 4.91(20) \\
            GR & 4.05(28) & 4.9(3) & 5.28(37) & 5.68(46) \\
            EM & 3.80(21) & 4.25(17) & 4.47(25) & 4.87(16) \\
            \bottomrule
        \end{tabular}
        \caption{Extrapolated crossings for the various values of the interaction $\Delta$ for the standard observables: participation entropy (PE), entanglement entropy (EE), gap ratio (GR), extreme magnetization (EM)}
        \label{tab:extrapolated}
    \end{table}

\begin{figure}[h!]
        \includegraphics[width =0.95\textwidth]{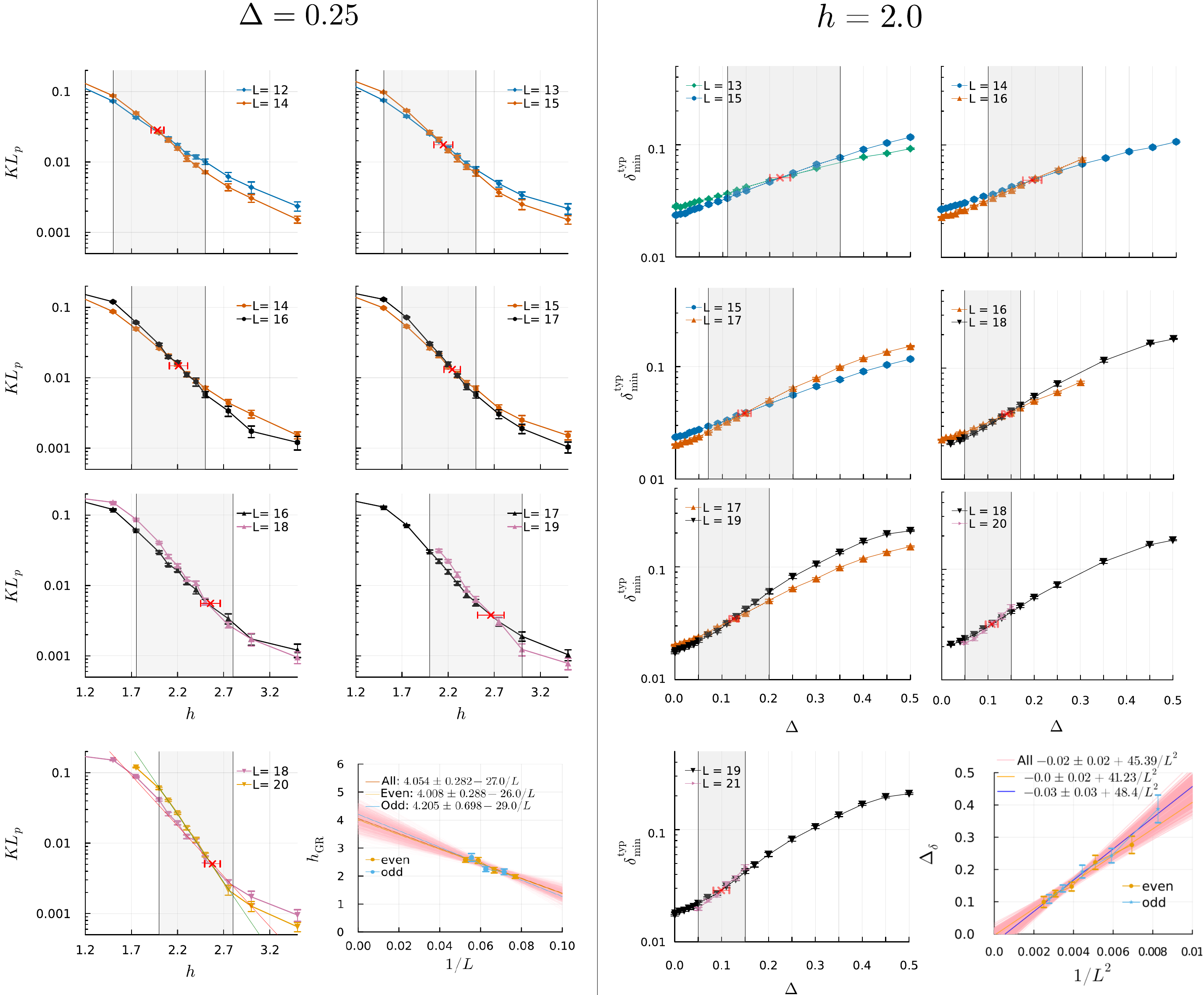}
        \caption{\label{fig:crossings} Extracting the critical line. Left: critical disorder strength at constant $\Delta = 0.25$ from the Kullback-Leibler divergence of the gap ratio. Right: critical interaction strength at constant disorder $h = 2$ from the extreme magnetization. 
        The gray area indicates the data which is used to extract a crossing. The crossing for a system size is denoted in red. The extracted crossings are then plotted against $1/L$ or $1/L^2$  ($L$ begin $L_{\rm{avg}}$ in all cases). In those plots, the numerous lines indicate several fits performed to evaluate the errorbars on the extracted result.
        All the bootstrap procedures are performed with $n = 1000$ fits.}
    \end{figure}

    As an example of extracting the crossings, in Fig.~\ref{fig:crossings} we details the KL divergence of the gap ratio at $h = 2$. 
    For the extraction of the crossings at constant interactions, we present results at $\Delta = 0.25$ for the minimal deviation in Fig.~\ref{fig:crossings}.

\begin{figure}[h!]
        \includegraphics[width=0.77\textwidth]{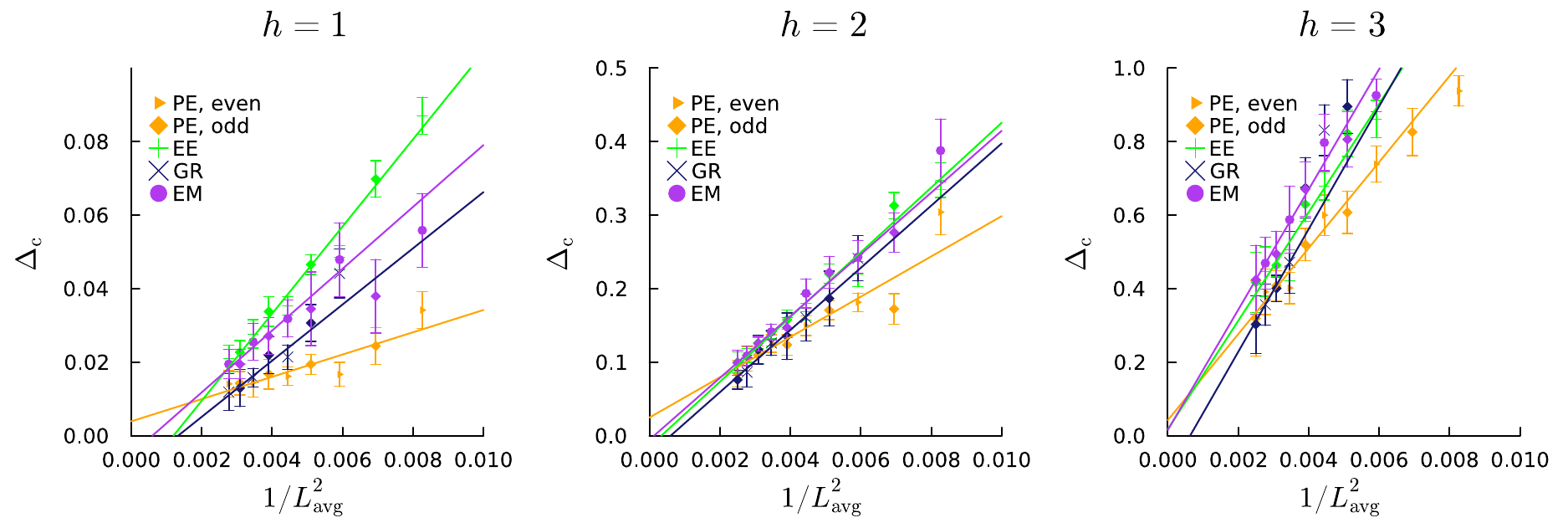}
        \caption{\label{fig:consth} Crossings for the four main observables and their drift with increasing system sizes for three different values of the field. Note the very differant ranges for the interaction axis. }
    \end{figure}

\begin{figure}
        \includegraphics[width=0.8\textwidth]{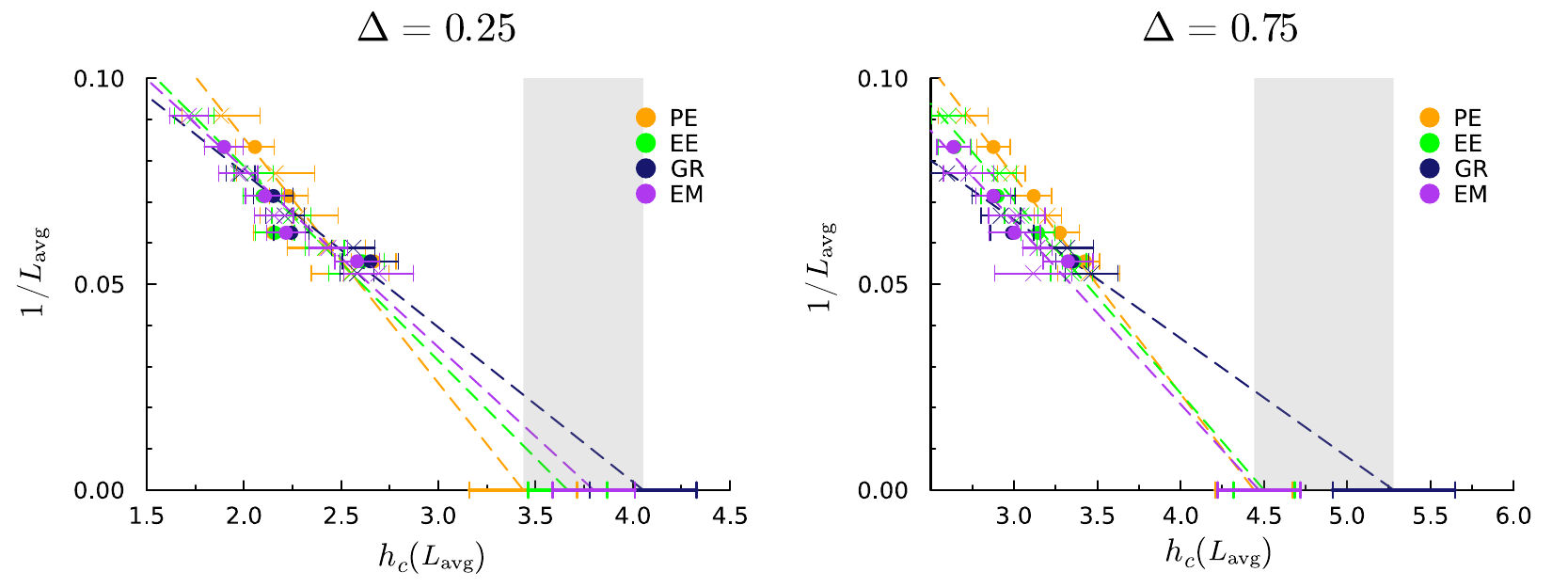}
        \caption{\label{fig:constd} Extrapolation as a function of $1/L_{\rm{avg}}$ of the crossings for the four observables.}
    \end{figure}

    In both cases, the procedure is the following for each pair of sizes : 
    \begin{enumerate}
        \item Select the gray area where the crossing seems to occur for each pair of sizes. 
        \item Bootstrap procedure to evaluate the crossings : Perform $n = 1000$ fits on the data (appropriate with the type of observable and the cut), where each fit is done removing 2 points  arbitrarily (if allowed by the number of points) and taking errors into account; obtain the crossing of these fits.
        \item The resulting crossings are shown as red crosses in Fig.~\ref{fig:crossings}. If the standard deviation is less than the spacing between data points, we take this spacing as the error. 
    \end{enumerate}
    Once the crossings are obtained with errors, we attempt to extract the critical disorder (resp. interaction) as $L\rightarrow \infty$. 
    For constant interaction (resp. disorder), we find that the fit that works best with most of the data is as a function of $1/L_{\mathrm{avg}}$ (resp. $1/L_{\mathrm{avg}}^2$), as depicted in Figs.~\ref{fig:consth} and~\ref{fig:constd}.
    The extrapolation in $1/L$ for constant interaction was already used in Ref.~\cite{sierant_polynomially_2020}.
    The behavior in $1/L_{\mathrm{avg}}^2$ is mostly a phenomenological observation. 
    

    The fits to extract the $L\rightarrow \infty$ limit are performed on even and odd sizes separately, and then on all data. In each case, the errors are evaluated from a bootstrap procedure. The final result is the best fit considering the largest sizes, but we check that the even and odd data fits are in agreement with this result within errors. 
    The thre panels of Fig.~\ref{fig:consth} show the overall drift of the crossings for $h = 1, 2$~and~$3$. The order of the \emph{extrapolated} critical value of the interaction from ergodic to MBL seems to be quite systematically
    PE, EM or EE, GR. This is in agreement with what we observe at constant interaction for the critical value of the disorder strength, where PE seems to be consistently underestimating the extent of the ergodic region as compared to the other observables. 
    
    Finally, Fig.~\ref{fig:constd} shows the extrapolation of $h_c(L)$ for the four observables PE, EE, EM and GR for $\Delta = 0.25$ and $\Delta = 0.75$ (similar to the insets of Fig.~3 in the main text). 
    \section{Simulations parameters}
    \label{sec:Params}
    The number of kept eigenstates (E)  and number of disorder realisations (S) are as follows for each size (Table~\ref{table:Samples}). In all cases, $N_{0.005}$ indicates the number of states that are within $\delta\epsilon = 0.005$ relative energy compared to $\epsilon = 0.5$, i.e. we consider $\leq 1\%$ of the spectrum. 
    \begin{table}[ht]
        \centering
        \begin{tabular}{c c c|c c c}
        \toprule
        $L$ & E & S & $L$ & E & S \\
        \midrule
        8  & 3               & 3500 & 14 \& 16 & $\min(60, N_{0.005})$  & $\geq$ 2500      \\ 
        9  & $\min(10, N_{0.005})$  & 3500  & 17 & $\min(60, N_{0.005})$  & $\geq$ 2500  \\ 
        10 & $\min(10, N_{0.005})$  & $\geq$ 2500  &  18 & $\min(60, N_{0.005})$  & $\geq$ 1500  \\ 
        11 & $\min(30, N_{0.005})$  & $\geq$ 2500  & 19 & $\min(60, N_{0.005})$  & $\geq$ 1000 \\ 
        12 & $\min(30, N_{0.005})$  & $\geq$ 2500  & 20 & $\min(60, N_{0.005})$  & $\geq$ 900 \\ 
        13 & $\min(40, N_{0.005})$  & $\geq$ 2500  & 21 & $\min(60, N_{0.005})$ & $\geq$ 750  \\
        \bottomrule
        \end{tabular}
        \caption{Details of the number of eigenstates kept and number of disorder samples for each size.}
        \label{table:Samples}
        \end{table}

\end{document}